# Nontrivial topological states in BaSn$_5$ superconductor probed by de Haas-van Alphen quantum oscillations

Lixuesong Han [1,†], Xianbiao Shi [2,3,†], Jinlong Jiao[4], Zhenhai Yu[1], Xia Wang[1,5], Na Yu[1,5], Zhiqiang Zou[1,5], Jie Ma[4], Weiwei Zhao[2,3], Wei Xia[1,6,*], Yanfeng Guo[1,6*]

[1]School of Physical Science and Technology, ShanghaiTech University, Shanghai 201210, China

[2]State Key Laboratory of Advanced Welding & Joining and Flexible Printed Electronics Technology Center, Harbin Institute of Technology, Shenzhen 518055, China

[3]Shenzhen Key Laboratory of Flexible Printed Electronics Techniology, Harbin Institute of Technology, Shenzhen, China

[4] Key Laboratory of Artificial Structures and Quantum Control (Ministry of Education), School of Physics and Astronomy, Shanghai Jiao Tong University, Shanghai 200240, China

[5]Analytical Instrumentation Center, School of Physical Science and Technology, ShanghaiTech University, Shanghai 201210, China

[6]ShanghaiTech Laboratory for Topological Physics, ShanghaiTech University, Shanghai 201210, China

[†]Lixuesong Han and Xianbiao Shi contributed equally to this work.

[*]Corresponding authors. xiawei2@shanghaitech.edu.cn, guoyf@shanghaitech.edu.cn

**We report herein the nontrivial topological states in an intrinsic type-II superconductor BaSn$_5$ ($T_c$ ~ 4.4 K) probed via measuring the magnetizations, specific heat, de Haas-van Alphen (dHvA) effect and performing first principles calculations. The first principles calculations reveal a topological nodal ring structure centering at the H point in the $k_z = \pi$** **plane of the Brillouin zone (BZ), which could be gapped by spin-orbit coupling (SOC), yielding rather small gaps below and above the Fermi level about 0.04 eV and 0.14 eV, respectively. The SOC also results in a pair of Dirac points along the Γ-A direction and located ~ 0.2 eV above the Fermi level. The analysis of the dHvA quantum oscillations supports the calculations by revealing nontrivial Berry phase originated from three hole and one electron pockets related to the bands forming the Dirac cones. Our study thus provides an excellent avenue for investigating the interplay between superconductivity and nontrivial topological states.**

The Majorana zero mode, which has been conceived to host potential applications for decoherence topological quantum computation owing to its virtue of non-Abelian statistics characteristic, [1–6] generally occurs at topological defects such as vortices, boundaries, domain walls, or edges of some effectively spinless superconducting systems with odd-parity triplet pairing symmetry, i.e. topological superconductors (TSCs).[2,5–9] Regarding such targets, the one-dimension (1D) $p$-wave and 2D $p_x \pm ip_y$ superconductors have been proposed.[9] Unfortunately, the pairing symmetry of the handful candidates, for example, the $Sr_2RuO_4$, $Cu_xBi_2Se_3$ and $Sr_xBi_2Se_3$, are still under hot debate.[10] Other ways to create equivalent $p + ip$ pairing superconductivity, including the construction of $Bi_2Te_3/NbSe_2$ heterostructure [11–13] and chemically doped superconductors, such as $FeTe_{0.55}Se_{0.45}$,[14–17] $(Li_{0.84}Fe_{0.16})OHFeSe$,[18,19] $CaKFe_4As_4$,[20] Co-doped $LiFeAs$,[21] etc., were reported to be effective to induce topological superconductivity. $FeTe_{0.55}Se_{0.45}$ among them is the most intensively studied one, which hosts both topological insulator (TI) and Dirac semimetal (DSM) states with the latter is slightly above the Fermi level ($E_F$). However, comparing with the heterostructures and doped superconductors, intrinsic TSCs are definitely more preferable for the study of the Majorana zero mode, because they are free of the disturbance by complicated interface physics and doping resulted detects or disorder. The intrinsic superconductors hosting nontrivial topological states therefore have been the targets due to that the interplay of nontrivial electronic band topology and superconductivity could serves as an excellent solid environment for the Majorana zero mode [9]. The examination of nontrivial topological states in superconductors is therefore a crucial intergradient for the discovery of intrinsic TSCs. Up to now, a series of intrinsic superconductors hosting nontrivial topological states have been investigated [22-29] and some of them were reported as TSCs. [29]

The $BaSn_5$ is an already known tin rich superconductor with a superconducting critical temperature $T_c$ of about 4.4 K. [30] It has a peculiar crystal structure as shown in Fig. 1(a), in which the Sn atoms are arranged in graphite-like honeycombs and two such honeycombs form the hexagonal prisms centered by Ba, which could be viewed as a variant of the well known $MgB_2$ family of superconductors. Previous studies suggested $BaSn_5$ as a *s*-wave multiband superconductor with isotropic upper critical field, while its electron-phonon coupling strength is debated.[30-32] More interestingly, van Hove singularity in the density of states arisen from the lone pairs was observed, which is usually discussed for high $T_c$ cuprate superconductors rather than such low $T_c$ intermetallic superconductor.[32] Despite of these interesting properties, $BaSn_5$ has been less studied, in particularly, in terms of its electronic band structure. Considering that several compounds with close compositions, such as CaSn, $CaSn_3$ and $BaSn_3$ superconductors,[24-26;33-35] host intriguing nontrivial topological states in the bulk band structures as well as nontrivial topological surface states, it is naturally an inspiration for the exploration of nontrivial topological states in $BaSn_5$. In this work, combining basic chracterizations on the superconducting properties, measurements of the dHvA effect and first principles calculations, we have demonstrated that the *s*-wave type-II $BaSn_5$ superconductor hosts intriguing multiple nontrivial topological states which highly resemble the case of $FeTe_{0.55}Se_{0.45}$.[14]

Single crystals of $BaSn_5$ were grown by using a Sn self-flux method.[31] Optical picture for a typical crystal is shown in Fig. 1(c). The crystallographic phase and quality of the crystal were examined on a Bruker D8 single crystal x-ray diffractometer with Mo K$_\alpha$ ($\lambda$ = 0.71073 Å) at 300 K. The result indicates a hexagonal structure with the space group of *P*6/*mmm* (No.191) and lattice parameters *a* = *b* = 5.37 Å, *c* = 7.097 Å, $\alpha$ = $\beta$ = 90°, $\gamma$ = 120°, which are nicely consistent with those values reported previously.[30] Seen in Figs. 1(a)-(b), the graphite-like Sn atoms form a two-dimensional slab of face-sharing hexagonal prisms which are doubly capped with Ba. The element-resolved energy dispersive spectroscopy (EDS) characterizations on more than ten different areas of several pieces of crystal revealed a good stoichiometry of 1 : 5, as shown in Fig. 1(c). The high quality of BaSn$_5$ single crystals could be evidenced by the perfect reciprocal space lattice of the single crystal X-ray diffraction without any impurities, which is shown in Figs. 1(d)-(f). Measurements of magnetizations and specific heat were on the magnetic property measurement system and physical property measurement system from Quantum Design, respectively.

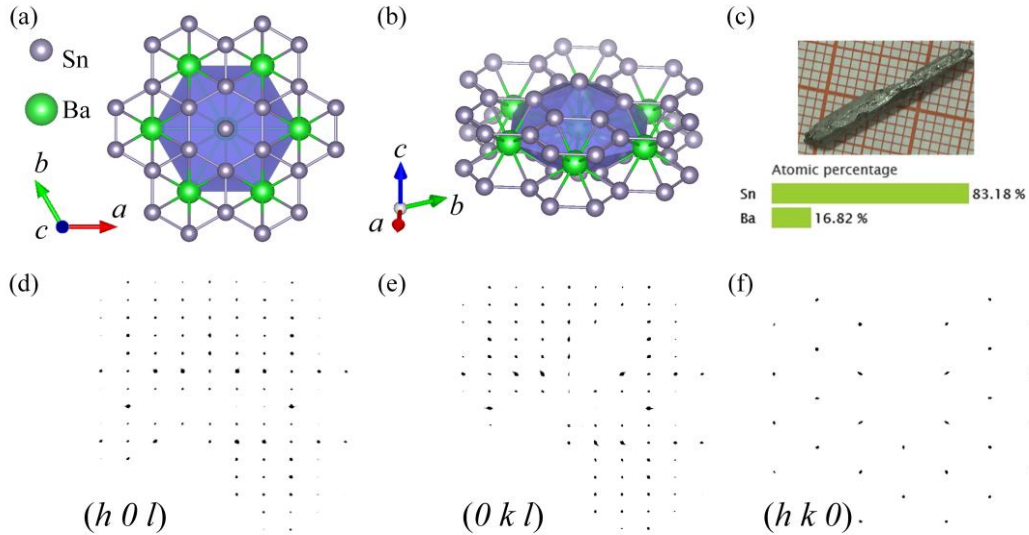

**Fig. 1.** (a)-(b) The schematic crystal structures of BaSn$_5$ viewed along different orientations. (c) The measured compositions by using EDS and an optical picture of a typical BaSn$_5$ crystal. (d)-(f) Diffraction patterns in the reciprocal space along the (*h 0 l*), (*0 k l*), and (*h k 0*) directions.

First principles calculations were performed within the framework of the projector augmented wave (PAW) method,[36,37] and employed the generalized gradient approximation (GGA)[38] with Perdew-Burke-Ernzerhof (PBE)[39] formula, as implemented in the Vienna *ab initio* Simulation Package (VASP).[40-42] For all calculations, the cutoff energy for the plane-wave basis was set to 500 eV, the Brillouin zone (BZ) sampling was done with a Γ-centered Monkhorst-Pack k-point mesh of size $11 \times 11 \times 8$, and the total energy difference criterion was defined as $10^{-8}$ eV for self-consistent convergence. The optimized structural parameters were used in the electronic structure calculations.

The temperature dependence of magnetizations with the magnetic field *B* of 10 Oe applied perpendicular to the (001) plane of the $BaSn_5$ crystal in the zero-field-cooling (ZFC) and field-cooling (FC) mode are shown in Fig. 2(a), showing a $T_c$ of ~ 4.4 K, consistent with that in previous reported literature.[31] Isothermal magnetizations measured at 2 K are depicted as an inset of Fig. 2(a), indicating a type-II superconductivity by the clear hysteresis. We would like to note that Sn is type-I superconductor with a $T_c$ ~ 3.6 K, which is not the case as $BaSn_5$. The bulk superconductivity is also demonstrated by the ~ 76.3% superconducting shielding volume fraction estimated from the ZFC data at 2 K.

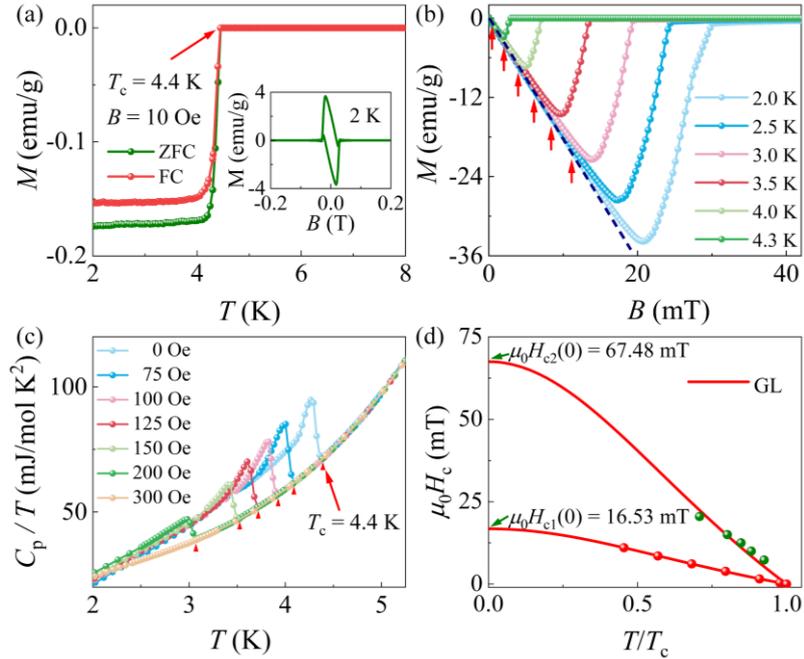

**Fig. 2.** (a) The temperature dependence of magnetization of BaSn$_5$ at 10 Oe. The inset shows the isothermal magnetization at 2 K. (b) Isothermal magnetizations measured at $T$ = 2 - 4.3 K with an interval of 0.5 K. (c) $C_p/T$ around $T_c$ with the magnetic field of 0 - 300 Oe. The arrows point out the positions for jump. (d) Upper (lower) critical field $H_{c2}$ ($H_{c1}$) versus the normalized temperature $T/T_c$. The solid lines denote the fitting results to $H_{c2}$ and $H_{c1}$ by using the G-L equation.

The bulk nature of the superconductivity is further confirmed by the clear jump in the specific heat $C_p(T)$ at $B$ = 0 T. Moreover, the $T_c$ observed in the $C_p(T)$ is ~ 4.4 K, nicely consistent with that derived from the magnetization characterization. To evaluate the lower and upper critical field, the temperature dependent isothermal magnetizations and magnetic field dependent low temperature specific heat of BaSn$_5$ are presented in Figs. 2(b) and 2(c), respectively. The lower critical magnetic fields $H_{c1}$ were determined by taking the magnetic field values at which the magnetizations start to deviate from the linear evolution, shown by the arrows in the inset to Fig. 2(b). This allows the extraction of the $H_{c1}(0)$ via fitting these values with employing the Ginsburg-Landau (GL) equation expressed as $H_{c1}(T) = H_{c1}(0)\frac{1-t^2}{1+t^2}$, where $t = T/T_c$ is the reduced temperature, giving $H_{c1}(0)$ = 16.53 mT. The superconducting jump in $C_p(T)$ exhibits an apparent suppression to low temperatures upon external magnetic field, which allows the construction of $H_{c2}$-$T$ diagram. By employing the GL equation expressed as $H_{c2}(T) = H_{c2}(0)\frac{1-t^2}{1+t^2}$ [34], the fitting result is shown by the solid line in Fig. 2(d). The upper critical magnetic field $H_{c2}(0)$ is 67.48 mT, comparable with that in previous work.[31]

To further understand the transport properties of BaSn$_5$, the electronic band structure of BaSn$_5$ was studied through the first principles calculations. As shown in Fig. 3(a), the band structure of BaSn$_5$ without considering SOC manifests a metallic character with several bands crossing E$_F$. As shown by the enlarged view in Fig. 3(b), two band-crossing points at -0.04 eV along the L-H path and at 0.14 eV along the H-A line are visible in the BZ. The fat band structure shows that the contribution to these crossing bands comes mainly from the Sn-5$p$ states, forming an inverted band structure at the H point. The couple of band-crossing points are not isolated but belong to a nodal ring centering the H point in the $k_z = \pi$ plane, as shown in Fig. 3(c). Hence, BaSn$_5$ superconductor could be classified as a topological metal with Dirac nodal ring structure when SOC is ignored.

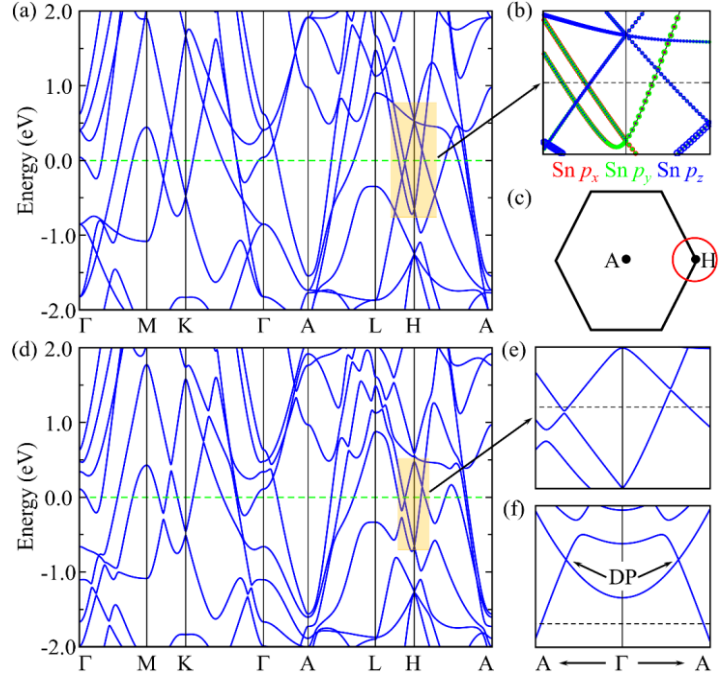

**Fig. 3.** (a) Electronic band structure of BaSn$_5$ without SOC considered. (b) Enlarged band structure along the L-H and H-A paths. The symbol size in (b) corresponds to the projected weight of Bloch states onto the Sn-$p_x$ (red), Sn-$p_y$ (green), and Sn-$p_z$ (blue) oribits. (c) Schematic illustration of the nodal ring centering the H point in the $k_z = \pi$ plane. (d) Electronic band structure of BaSn5 with SOC considered. (e) Same as (b) but with SOC included. (f) Enlarged band structure along the A-Γ path, showing a pair of Dirac points.

When the SOC is considered, as presented in Fig. 3(d), the nodal line of BaSn$_5$ is gapped, showing a rather small gap with the maximum size of about 20 meV along the L-H path of the Brillouin zone, which is in fact much smaller than the energy dispersion of the nodal ring (about 180 meV). Moreover, the SOC induces a pair of Dirac points (DPs) along the Γ-A direction, as shown in Fig. 3(f). Though the DPs locate about 0.2 eV above the $E_F$, they can contribute to the transport behavior due to the Dirac cone bands cross the $E_F$. Hence, BaSn$_5$ superconductor hosts both DSM and TI states when SOC is considered.

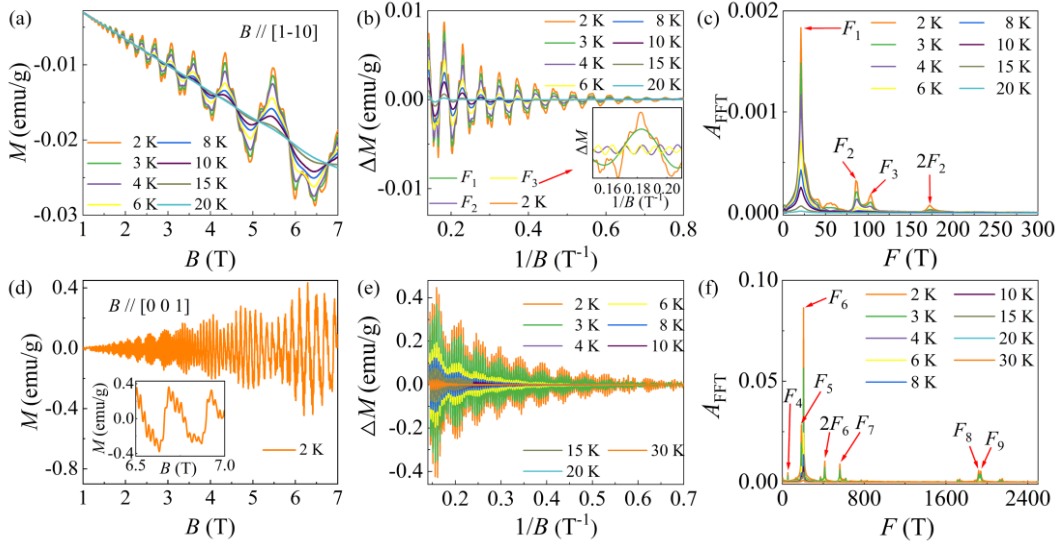

**Fig. 4.** (a) and (d) Magnetizations of BaSn$_5$ vs. $B$ for $B$//[1-10] and $B$//[001] respectively at various temperatures. (b) and (e) The quantum oscillations for $B$//[1-10] and $B$//[001] vs. $1/B$ respectively at various temperatures. The inset in (b) shows the three filtered oscillatory parts of $\Delta M$. The orange line represents the raw dHvA oscillatory signal of BaSn$_5$ at 2 K. (c) and (f) The Fast Fourier Transforms (FFT) spectra of $\Delta M$ at various temperatures.

To probe the Fermi surface (FS) of BaSn$_5$, the dHvA effect was measured. As shown in Figs. 4(a) and 4(d), we measured the isothermal magnetization of BaSn$_5$ up to 7 T with $B$//[1-10] and $B$//[001] at various temperatures. As shown in Figs. 4(b) and 4(e), after subtracting the background, the magnetizations $\Delta M$ (= $M$ - $M_{\text{background}}$) display striking oscillations between 2 - 30 K, which could be well described by the Lifshitz-Kosevich (L-K) formula [43]

$$\Delta M \propto B^\lambda R_T R_D R_S \sin\left[2\pi\left(\frac{F}{B} - \gamma - \delta\right)\right],$$

where $R_T = 2\pi^2 k_B T/\hbar\omega_c / \sinh(2\pi^2 k_B T/\hbar\omega_c)$ with $\omega_c = eB/m^*$ being the cyclotron frequency and $m^*$ denoting the effective cyclotron mass.

$R_D = \exp(-2\pi^2 k_B T_D/\hbar\omega_c)$ with $T_D$ is the Dingle temperature and $R_S = \cos(\pi g m^*/2m_e)$. Index $\lambda$ is determined by the dimensions, with the $\lambda$ denoting 1/2 in three-dimensional (3D) case and 0 in 2D case [44]. Phase factor $(-\gamma - \delta)$ is utilized to describe the oscillation part, where $\gamma = 1/2 - \Phi_B/2\pi$ with $\Phi_B$ being the Berry phase. The phase shift is determined by the dimensionality of the FS and takes a value of 0 for 2D and ±1/8 ("-"for the electron-like pocket and the "+"for the hole-like pocket) for the 3D case. The first-order differential of $M$ vs. $1/B$ allows the analysis of the fast Fourier transformation (FFT), as shown in Figs. 4(c) and 4(f). When $B$ // [1-10] and $B$ // [001], multiple fundamental frequencies were obtained, as shown in Figs. 4(c) and 4(f), respectively. For more accurate fitting, band-pass filtering was performed to separate the low and high frequency components. The thus obtained fundamental frequencies are labeled as $F_1$ (21 T), $F_2$ (86.4 T), $F_3$ (103.3 T), $F_4$ (53.3 T), $F_5$ (185 T), $F_6$ (207 T), $F_7$ (561 T). $2F_2$ and $2F_6$ are the harmonic value that indicates the same FS for $F_2$ and $F_6$. The multiple fundamental frequencies imply that several Fermi pockets are across or close to $E_F$. The areas of Fermi pockets $A_F$ can be calculated by the Onsager relationship $F = A_F(\varphi/2\pi^2)$. The effective mass $m^*$ can be obtained by fitting the temperature-dependent oscillation amplitude to the thermal resistance factor $R_T$, shown in Figs. 5(b) and 5(e). As shown in Figs. 5(c) and 5(f), via fitting to the field-dependent amplitude of the quantum oscillations at 2 K, the obtained Dingle temperatures $T_D$ are evaluated for the fundamental frequencies $F_1$ - $F_7$, respectively. The other parameters are summarized in Table 1.

The Landau level (LL) fan diagrams were established to test the Berry phase of $BaSn_5$ accumulated along with the cyclotron orbit. The LL phase diagram for $F_1$ - $F_3$ and $F_4$ - $F_7$ are shown in Figs. 5(a) and 5(d), respectively, where the valley positions of $\Delta M$ correspond to the Landau indices of $n = N - 1/4$, and the peak positions of $\Delta M$ correspond to the Landau indices of $n = N + 1/2 - 1/4$.[44] The intercepts of the linear fitting are 0.379, 0.394, 0.425, 0.018, 0.597, 0.245, and 0.101, corresponding to fundamental frequencies $F_1$ - $F_7$. The data ambiguously unveil the 3D FS of $BaSn_5$.

As shown in Fig. 6(b), these angle-dependent frequencies $F_1$ - $F_7$ are consistent with the calculated values for bands 1, bands 2, bands 3, bands 4 illustrated by the Fig. 6(c). The angle dependent dHvA oscillations could provide further information about the shape of the FS. The oscillations amplitudes versus $1/B$ at various $\theta$ and at 2 K are shown in Fig. 6(a). For BaSn$_5$, band 1(red) contributes hole pockets, while bands 2 (blue), 3 (orange), 4 (olive green) contribute to electron pockets. With a careful comparison between experimental and theoretical values, $F_4$ and $F_1$ are assigned to electron pockets pointed by band 4, while $F_5$, $F_2$, $F_6$, $F_3$, and $F_7$ are assigned to hole pockets pointed by band 1. Thus, the Berry phases of $F_1$ - $F_7$ are $2(0.379 - 0.125)\pi$, $2(0.394 + 0.125)\pi$, $2(0.425 + 0.125)\pi$, $2(0.018 - 0.125)\pi$, $2(0.596 + 0.125)\pi$, $2(0.245 + 0.125)\pi$ and $2(0.101 + 0.125)\pi$, respectively. The Berry phases for $F_2$, $F_3$, $F_5$, $F_6$ are close to $\pi$, indicating nontrivial topological nature of band 1, which is consistent with the first principle calculations.

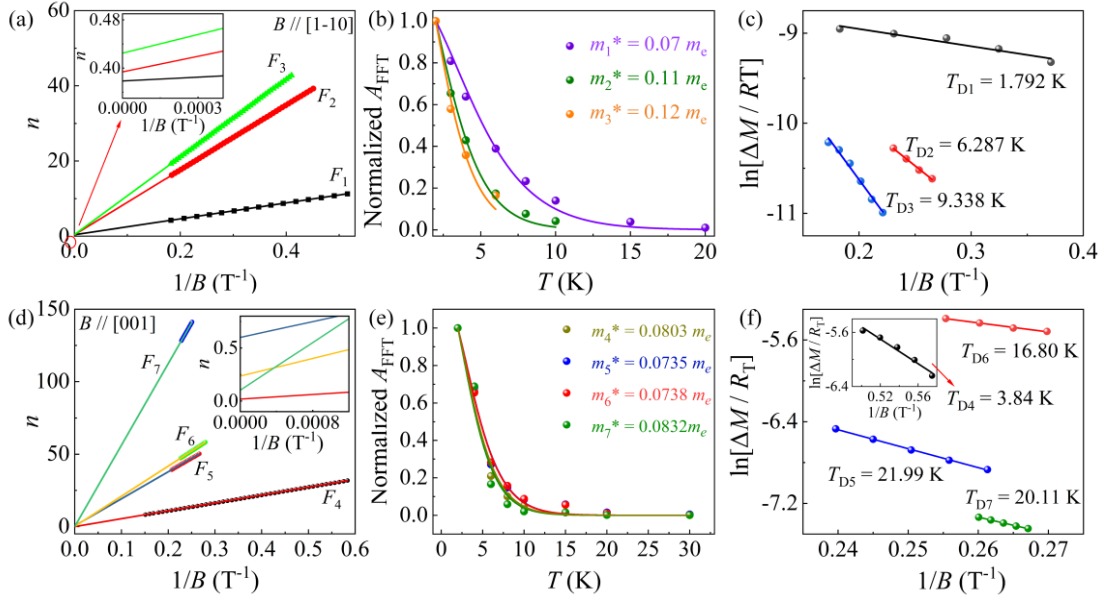

**Fig. 5.** (a) and (d) Landau indexes $N - 1/4$ plotted against $1/B$ at 2 K for $B//[1\text{-}10]$ and $B//[001]$, respectively. The inset in each enlarges the intercepts of the fitting. (b) and (e) Temperature dependence of relative FFT amplitudes of the oscillations for $B//[1\text{-}10]$ and $B//c$, respectively. (c) and (f) Dingle plots of dHvA oscillations at 2 K for $B//[1\text{-}10]$ and $B//c$, respectively.

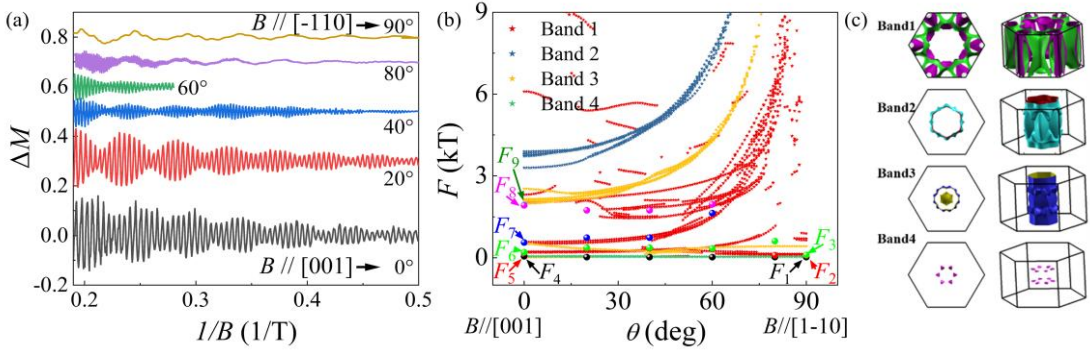

**Fig. 6.** (a) The FFT spectra of $\Delta M$ at different $\theta$ from $B//[001]$ to $B//[1\text{-}10]$. (b) The comparison between experimental and theoretical values of the fundamental frequencies with varied $\theta$. (c) The calculated Fermi surface, in which band 1 contributes hole pocket, while bands 2, 3, and 4 contribute to electron pockets.

Table 1. Parameters derived from dHvA oscillations for BaSn$_5$, wehere $k_F$ is the Fermi wave vector, $v_F$ denotes the Fermi velocity, $\tau_Q$ is the relaxation time and $\Phi_B$ is the Berry phase.

|  | $F$ (T) | $A$ (nm$^{-2}$) | $K_F$ (nm$^{-1}$) | $v_F$ (m/s) | $m^*/m_e$ | $T_D$ (K) | $\tau_Q$ (s) | Berry phase |
|---|---|---|---|---|---|---|---|---|
| $F_1$ | 21 | 0.200 | 0.252 | $4.3 \times 10^5$ | 0.07 | 1.792 | $6.78 \times 10^{-13}$ | $0.508\,\pi$ |
| $F_2$ | 86.4 | 0.822 | 0.512 | $5.5 \times 10^5$ | 0.11 | 6.287 | $1.93 \times 10^{-13}$ | $1.038\,\pi$ |
| $F_3$ | 103.3 | 0.985 | 0.560 | $5.2 \times 10^5$ | 0.12 | 9.338 | $1.30 \times 10^{-13}$ | $1.100\,\pi$ |
| $F_4$ | 53.3 | 0.508 | 0.402 | $5.8 \times 10^5$ | 0.08 | 3.84 | $3.17 \times 10^{-13}$ | $-0.214\,\pi$ |
| $F_5$ | 185 | 1.764 | 0.750 | $1.05 \times 10^6$ | 0.07 | 21.99 | $5.5 \times 10^{-14}$ | $1.440\,\pi$ |
| $F_6$ | 207 | 1.973 | 0.793 | $1.09 \times 10^6$ | 0.07 | 16.80 | $7.2 \times 10^{-14}$ | $0.740\,\pi$ |
| $F_7$ | 561 | 5.349 | 1.305 | $1.59 \times 10^6$ | 0.08 | 20.11 | $6.0 \cdot 10^{-14}$ | $0.452\,\pi$ |

To summarize, we have demonstrated nontrivial topological states in the BaSn$_5$ superconductor. The chracterizations by magnetizations and specific heat measurements unveil the type-II superconductor nature. The dHvA quantum oscillations and the first principles calculations suggest topological nodal-line structure in BaSn$_5$, which can be gapped by taking the SOC into account and, alternatively, a pair of DPs appear. Interestingly, $BaSn_5$ hosts both TI and DSM states in the electronic band structure, highly resembling the case in $FeTe_{0.55}Se_{0.45}$ and Co-doped LiFeAs.[14,21] If the FS formed by the Dirac Fermi arc can corporate with the bulk superconductivity, it is promising to form an effective proximity effect in momentum space, which could serve the solid environment for the Majorana zero mode. Comparing with $BaSn_3$ and $CaSn_3$,[24,35] the DPs are more close to the $E_F$ in $BaSn_5$, which thus provides an excellent example for the study on the correlation between nontrivial topological states and conventional superconductivity.

The authors acknowledge the supported by the National Natural Science Foundation of China (Grant No. 92065201, 11774223 and U2032213), the open project of Key Laboratory of Artificial Structures and Quantum Control (Ministry of Education), Shanghai Jiao Tong University (Grant No. 2020-04). W.W.Z. is supported by the Shenzhen Peacock Team Plan (Grant No. KQTD20170809110344233) and Bureau of Industry and Information Technology of Shenzhen through the Graphene Manufacturing Innovation Center (Grant No. 201901161514). The authors thank the support from Analytical Instrumentation Center (#SPST-AIC10112914), SPST, ShanghaiTech University.